\documentclass[final,3p,times,twocolumn]{elsarticle}

\usepackage{graphicx,bm,epsf,float,amssymb,caption}

\usepackage{url} 
\usepackage{subfigure}
\usepackage{lineno}
\usepackage{color}

\biboptions{sort&compress}

\DeclareGraphicsExtensions{.pdf,.png,.jpg}

\journal{Nuclear Instruments and Methods A}

\begin{document}

\begin{frontmatter}
\title{PMT signal increase using a wavelength shifting paint}

\author[mit]{K.~Allada}
\author[eljen]{Ch. Hurlbut}
\author[mit]{L.~Ou}
\author[mit]{B.~Schmookler}
\author[yerevan]{A.~Shahinyan}
\author[jlab]{B.~Wojtsekhowski \corref{cor}}\ead{bogdanw@jlab.org} 

\address[mit]{Massachusetts Institute of Technology, Cambridge, MA 02139}
\address[eljen] {Eljen Technology, Sweetwater, TX 79556}
\address[yerevan] {Yerevan Physics Institute, Armenia}
\address[jlab]{Thomas Jefferson National Accelerator Facility, Newport News, VA 23606}

\cortext[cor]{Corresponding Author. Tel: +1 757 269 7191}

\begin{abstract}
We report a 1.65 times increase of the PMT signal and a simple procedure of application of a new wavelength shifting (WLS) paint for PMTs with non-UV-transparent windows. Samples of four different WLS paints, made from hydrocarbon polymers and organic fluors, were tested on a 5-inch PMT (ET 9390KB) using Cherenkov radiation produced in fused silica disks by $^{106}$Ru electrons on a `table-top' setup. The best performing paint was employed on two different types of 5-inch PMTs (ET 9390KB and XP4572B), installed in atmospheric pressure CO$_2$ gas Cherenkov detectors, and tested using GeV electrons.
\end{abstract}

\end{frontmatter}

\section{Introduction}
Detection of Cherenkov radiation produced by a fast charged particle passing through a medium is one of the common ways to identify the type of particle in nuclear and particle physics experiments. A typical Cherenkov counter consists of a radiator, followed by a mirror, from which the photons bounce off for detection in a PMT. The photon transmission through the radiator, the mirror reflectivity, and the quantum efficiency of the PMT vary with the wavelength of  the photons striking the PMT's window, while the intensity of the Cherenkov light is inversely proportional to the square of the photon's wavelength. The corrections for the gas transparency and mirror reflectivity are relatively small in the region of wavelengths above 200~nm. PMTs which have ultraviolet (UV) transparent windows, such as a quartz window, can detect this radiation more efficiently but are usually very costly. A less expensive and well known option is to use PMTs with non-UV-transparent windows (e.g. borosilicate) and coat them with wavelength shifting material~\cite{garwin, baillon, paneque}.

\begin{figure}[t]
\centering
\includegraphics[width=0.5\textwidth]{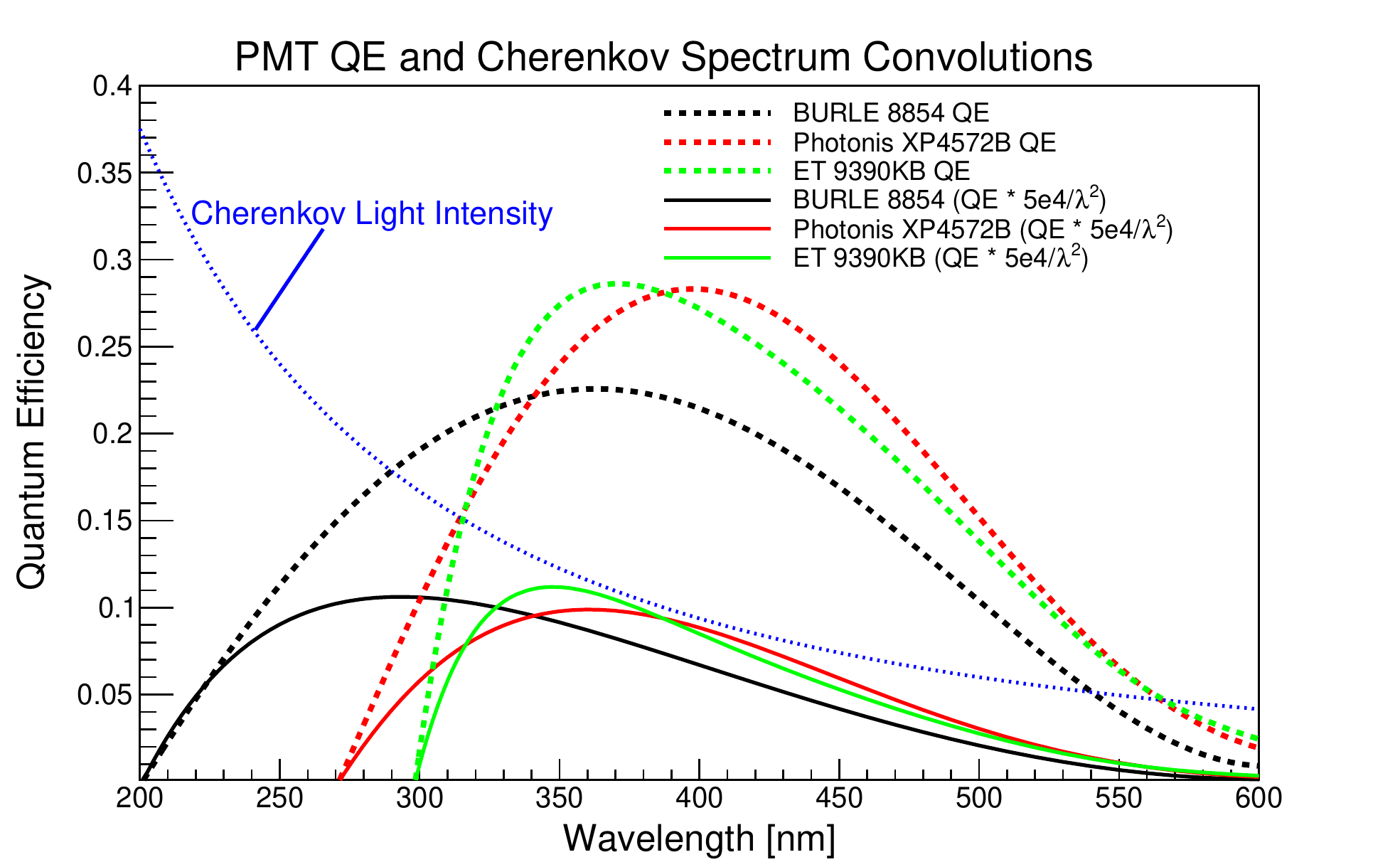}
\caption{The spectrum of Cherenkov radiation, PMT quantum efficiency and the convolution for three types of PMTs. The result (solid lines) leads to approximately 22, 17, and 16 photoelectrons for the \textit{Burle}, \textit{Photonis}, and \textit{ET} tubes, respectively.} 
\label{fig:PMT_QE}
\end{figure}

The effect of the difference in quantum efficiency for different types of PMTs on the ability to detect Cherenkov radiation is illustrated in Fig.~\ref{fig:PMT_QE}. The number of photoelectrons that a PMT would generate per event is given by the integral over all wavelengths of the product of the number of photons striking the PMT and the PMT's quantum efficiency. This product is shown by the solid curves in the figure, with the number of photons corresponding to the 1.2~m long atmospheric pressure CO$_2$ Gas Cherenkov detectors in the HRS spectrometers of Hall A at Jefferson Lab.

The application of wavelength shifting (WLS) material, which absorbs light in the ultraviolet spectrum and emits it in the visible range, to the surface of a PMT with a non-UV-transparent window can allow such a PMT to detect Cherenkov radiation more effectively. WLS materials are generally prepared from organic compounds. Some of these materials, such as p-terphenyl (pTP), are applied to surfaces using vapor deposition~\cite{garwin, baillon} or dip coating~\cite{paneque, CBM}.
Different WLS materials have been tested on non-UV-transparent PMTs in the past, resulting in a variety of gain improvements~\cite{garwin, baillon, paneque, boccone, hohne, TUNKA}. (WLS paints have also been used with UV-transparent PMTs, but with relatively modest enhancement on the level of 15-20\%~\cite{CBM}.) We report here a study of the signal gain for non-UV-transparent PMTs using the WLS paints developed by Eljen Technology~\cite{eljen}.

\begin{figure}[t]
\centering
\includegraphics[width=0.45\textwidth]{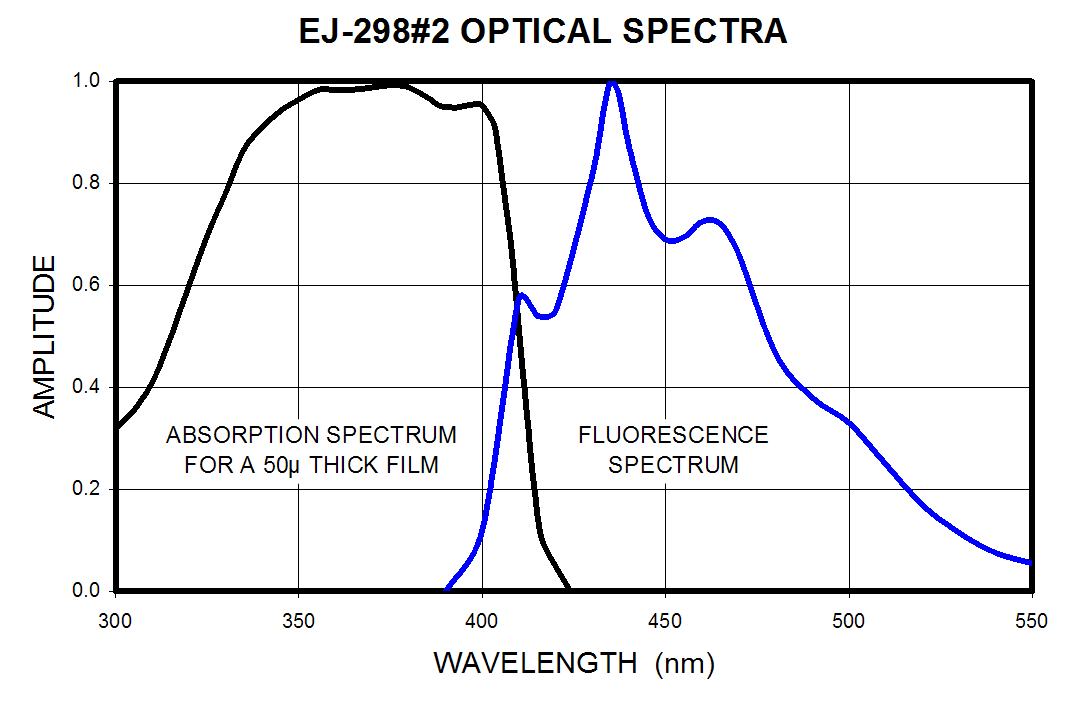}
\caption{Absorption and fluorescence spectra for the WLS paint EJ-298\#2 from Ref.~\cite{eljen}. The $y$-axis displays the fraction of light absorbed at a given wavelength; the remainder is either scattered or transmitted.} 
\label{fig:paint_spectrum}
\end{figure}

The WLS paints we tested contain a hydrocarbon polymer, which is the binder of the dried paint, and organic fluors dissolved in a xylene or an MEK (Methyl Ethyl Ketone) base~\cite{eljen}. While the paint remains a fluid, the solvent is the dominant component; the binder is the main component once the paint dries. The typical absorption and fluorescence spectra of a WLS sample (EJ-298\#2) developed by Eljen Technology are shown in Fig.~\ref{fig:paint_spectrum}.  

We present here briefly the analysis of the number of photoelectrons from a PMT's charge spectrum. This is followed by a description of a simple technique for applying the WLS paint to the surface of a PMT window and a discussion of a `table-top' setup to test various WLS paints. Finally, the results from in-beam tests with GeV electrons are given.

\section{Number of Photoelectrons from Charge Spectrum}
In order to obtain information on the number of photoelectrons detected by a PMT, the signal charge spectrum must be analyzed. The key assumption that the electron cascade that follows the emission of photoelectrons from the photocathode proceeds independently for each photoelectron works well for a 5-inch PMT up to a gain of 10$^8$. Therefore, the charge spectrum of an event with $n$ photoelectrons is a statistical sum of $n$ signals, each of which has the charge spectrum of the single-photoelectron (S-PE)~\cite{TN, NIM-D}. Naturally, the mean of the $n^{th}$-photoelectron spectrum is $n$ times larger than the mean of the S-PE spectrum, mS-PE. The S-PE spectrum and mS-PE are determined solely by the properties of the PMT. If the mS-PE value is known, the charge spectrum can be easily converted into the photoelectrons detected.

\begin{figure}[thb]
\centering
\includegraphics[width=0.5\textwidth]{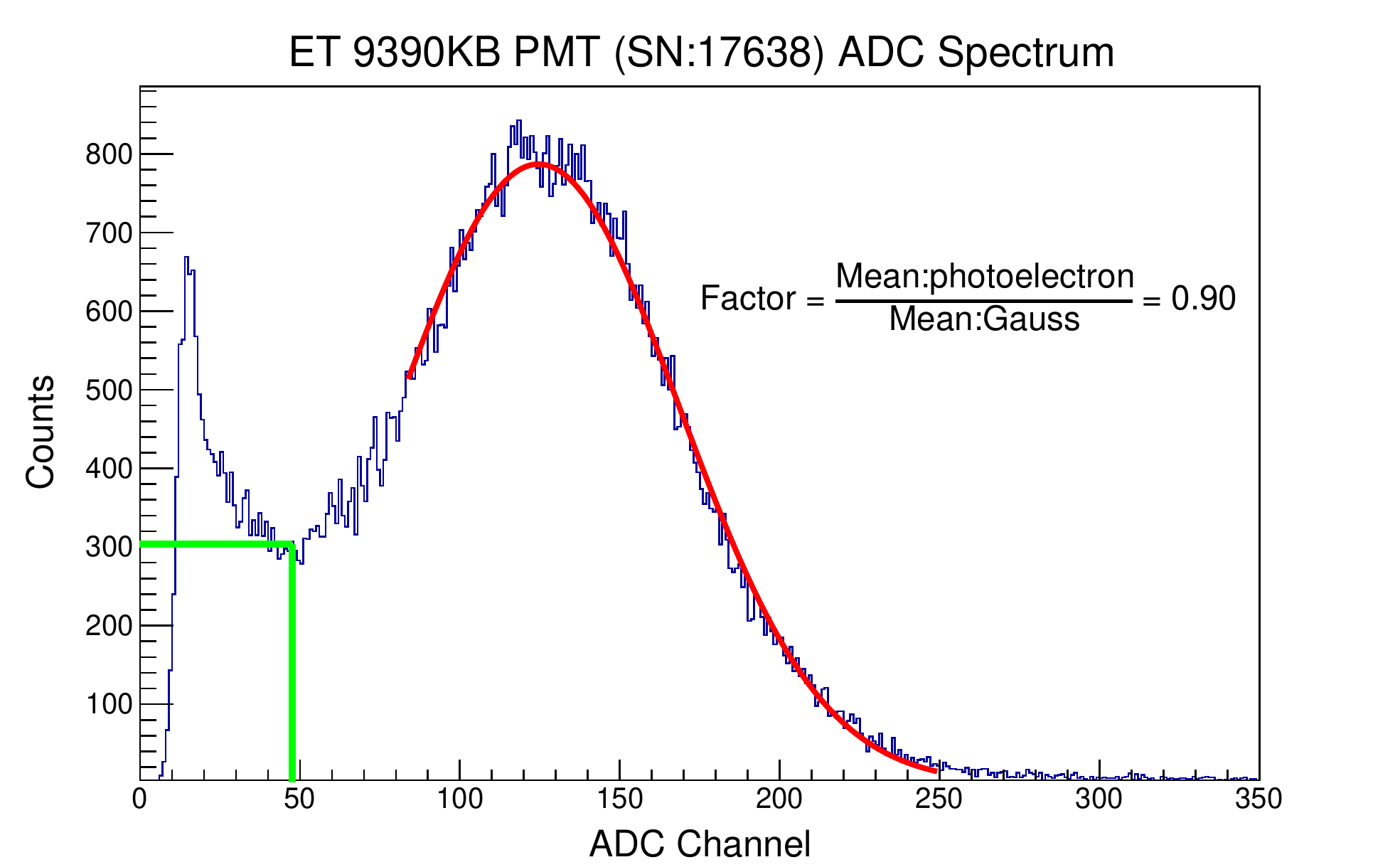} 
\caption{Self-triggered charge spectrum for an {ET 9390KB} PMT. The peak area was fit with a Gaussian function. The location of the plateau can be seen by the green lines.}  
\label{fig:factor_17638}
\end{figure}

An approximate method was developed to determine the mS-PE for a given PMT. The PMT was placed in a dark enclosure, and its charge spectrum was examined. Since the PMT served as its own trigger,  this `dark-current' charge spectrum consisted almost entirely of events where a single photoelectron was produced. Fig.~\ref{fig:factor_17638} shows this charge spectrum for an {ET 9390KB} PMT at a certain high-voltage setting. The spectrum has a peak, usually called a single photoelectron peak, which is fitted with a Gaussian. Proceeding away from the peak towards lower charge, the spectrum reaches a minimum and then increases until it is abruptly cut off, that point being determined by the trigger threshold. We would like to note that it is incorrect to attribute all low-charge events in the spectrum below the peak to the electronic noise and treat the S-PE spectrum as a pure Gaussian distribution (see the full analysis of this subject in Refs.~\cite{TN, NIM-D}). 

The spectrum was modeled as following a Gaussian distribution until the local minimum at low charge. The low-charge events are thought to come from a combination of photoelectrons elastically scattering off the first dynode, thermoionic emission from potential photocathode material on the inside walls of the PMT, and electronic noise. Only the portion of the low-amplitude region attributed to elastic scattering should be included for the calculation of photoelectron yield. The elastic scattering contribution is modeled as following a plateau from the minimum at low charge down to zero (where the pedestal would be located)~\cite{TN}. The ratio of the mean of this full spectrum compared to the mean of the Gaussian peak was calculated. It was found to be $R_{ET9390KB} = 0.90$. This ratio holds for all PMTs of this type for any applied voltage, so the mS-PE spectrum can be determined by fitting the Gaussian portion of the spectrum and multiplying the mean of the fit by the known ratio, $0.90$. The same procedure was performed for {Photonis XP4572B} PMTs. The ratio for these PMTs was found to be $R_{Photonis  XP4572B} = 0.95$. 

\section{Application of WLS Paint to the PMTs}
Four different paints developed by Eljen Technology were applied to ET 9390KB PMTs and tested in the `table-top' setup described in Sec.~\ref{sec:TTS}. 
The first paint tested, EJ-298\#2, is composed of the hydrocarbon polymer polyvinyltoluene (PVT) and organic fluors in a xylene base~\cite{eljen}. This paint was initially applied to the PMTs using a foam brush. A single application of the paint resulted in a layer thickness of 25-75~$\mu$m, and additional coatings could be applied to increase the layer thickness. The paint was tested in the `table-top' setup described in the next section with thicknesses ranging from 25-200~$\mu$m, and no dependence on the thickness was seen in this range. From these results and from calculations of the optical densities of the fluors employed in the different paints, it was decided to keep the paint thickness between 25 and 100~$\mu$m.

The other three paints tested (EJ-299-31(A,C,E)) employ the same hydrocarbon polymer polymethyl methacrylate (PMMA) in the MEK base, but contain different fluorescent additives~\cite{eljen}. The paints EJ-299-31A and EJ-299-31C each contain a single fluor with an emission maximum near 430~nm and 370~nm, respectively. The paint EJ-299-31E combines the above two fluors and contains an additive to retard the drying rate. Attempts to apply these paints with a brush were not very successful and resulted in large non-uniformities in the applied layer.

\begin{figure}[htb]
\includegraphics[width=0.25\textwidth]{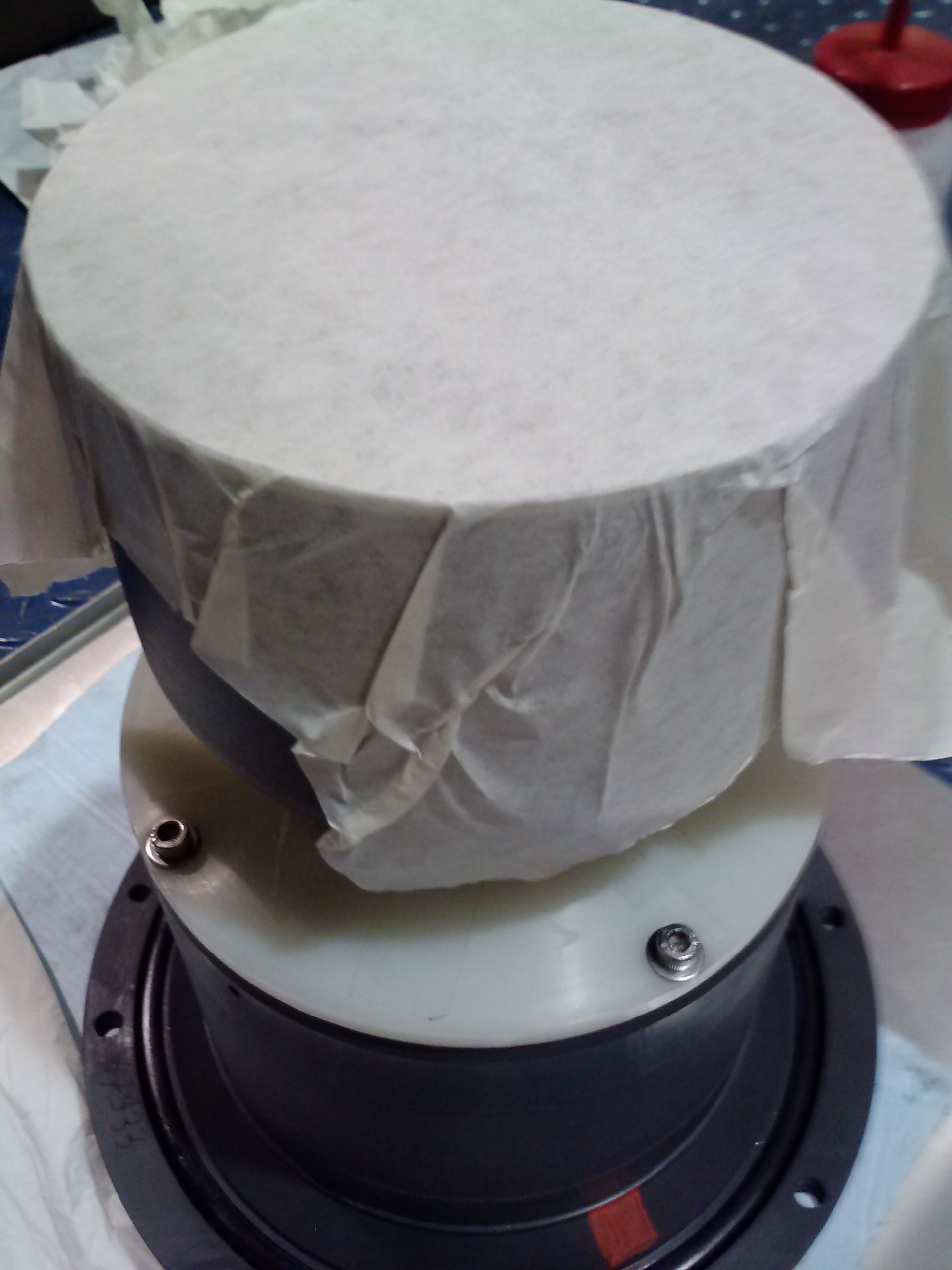}
\includegraphics[width=0.22\textwidth]{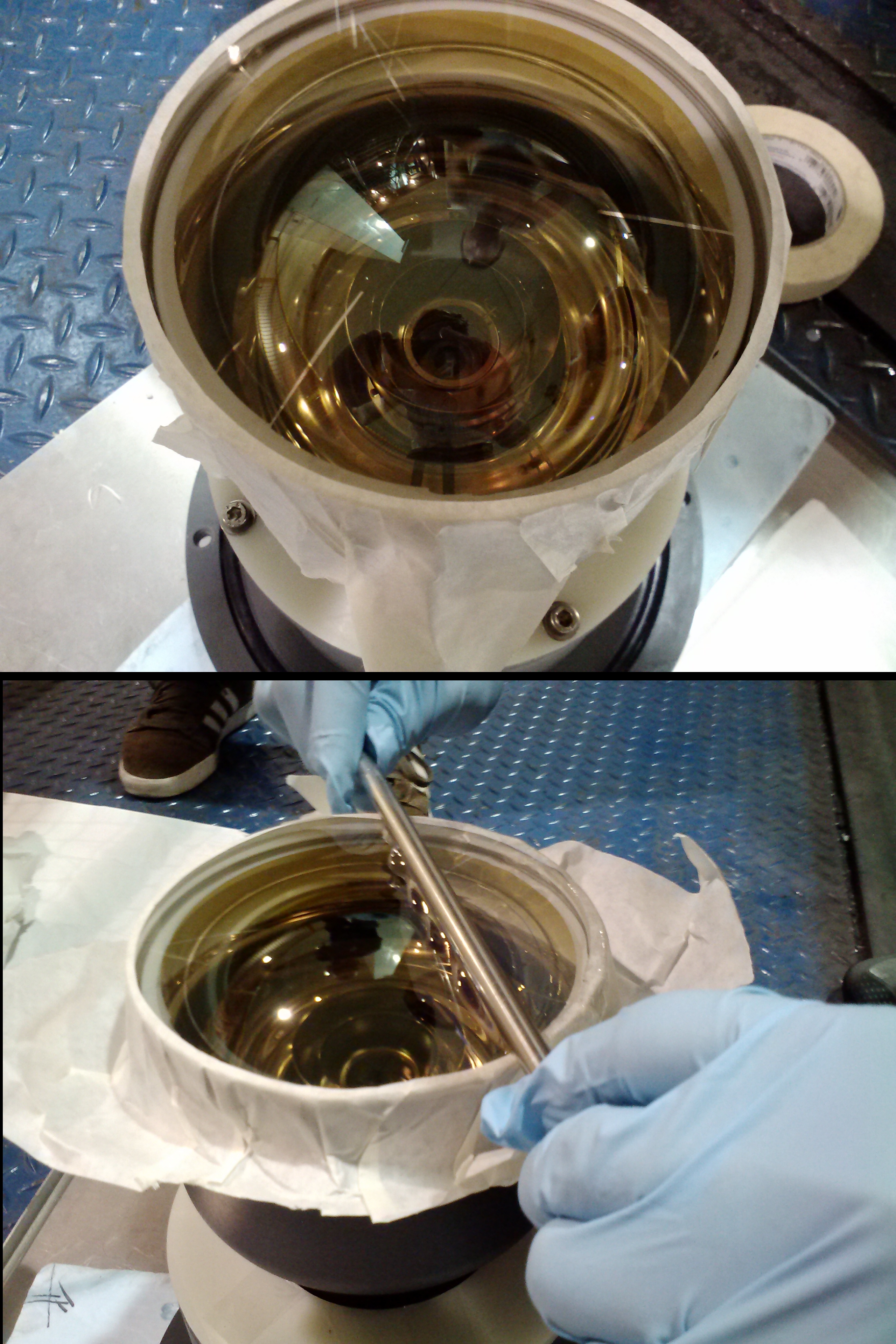}
\caption{The adhesive film applied to the surface of the PMT (Left). Ring of the adhesive film remaining after the central part
was removed using a circular template (Top-right). Application of WLS paint with metal rod drawn over the surface of the PMT (Bottom-right).} 
\label{fig:Adhesive_Full}
\end{figure}

A better technique, which allows for uniform application and better
control of the WLS thickness, is the `draw-bar' technique. In this method a metal rod is used to spread paint over the surface of the PMT while being kept at a fixed height above the surface. The PMT window was first cleaned with isopropanol and installed upward horizontally in a fixture. A thin adhesive film of thickness $\approx90~\mu$m was then applied to the surface of the PMT (Fig.~\ref{fig:Adhesive_Full}). A circular template of radius slightly smaller than the PMT radius was used to remove the
central part of the film from the PMT surface. 

\begin{figure}
\centering
\includegraphics[width=1.5in]{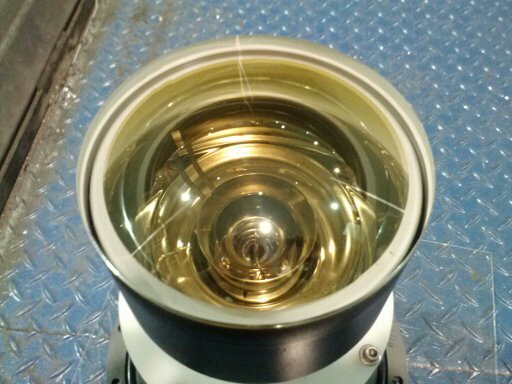}
\includegraphics[width=1.5in]{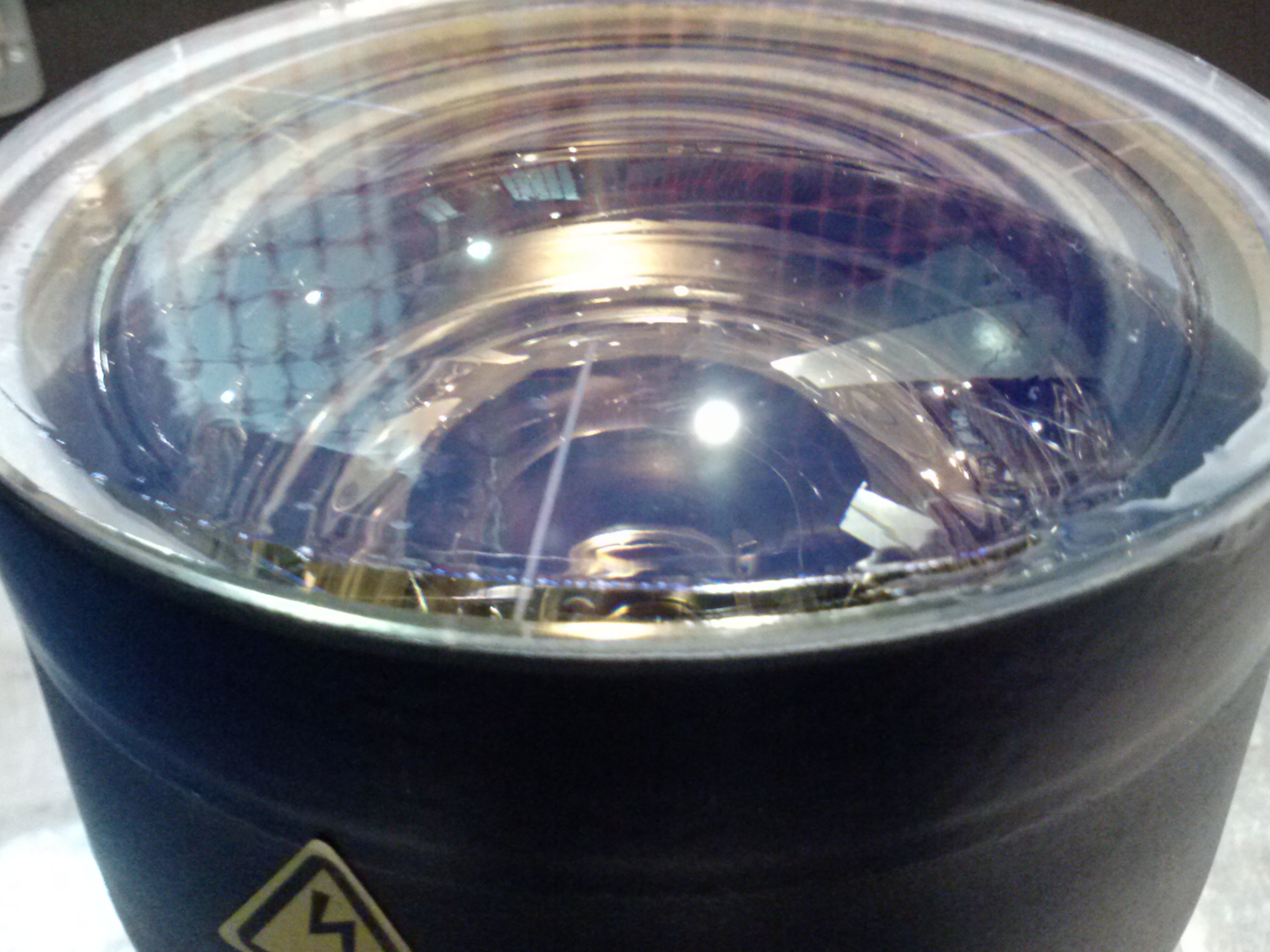}
\caption{Picture of {Photonis 4572B} PMT without WLS applied (Left). Same PMT after WLS paint applied (Right).}  
\label{fig:Photonis_b&a}
\end{figure}

A small amount of paint (approximately 2mL) was poured onto the PMT surface using a syringe, and the rod was drawn across the surface while resting on the adhesive film (Fig.~\ref{fig:Adhesive_Full}). 
The paint was allowed to dry for half an hour before the PMT was removed from the fixture.

The thickness of the WLS layer after it dried was found to be of 30~$\mu$m (1/3 of the wet layer thickness), but this dry layer thickness varies by about 10-15~$\mu$m across the surface of a specific PMT. A factor here is the flatness of the PMT surface, which was found to vary by a maximum of 50-75~$\mu$m over the full 5-inch diameter of the tube.

Using the `draw-bar` technique, we were able to successfully apply the paints EJ-298\#2 and EJ-299-31E to the PMT surface. The paints EJ-299-31A and EJ-299-31C, however, would dry too quickly, and the WLS layer would be significantly non-uniform.
A before-and-after photo of a {Photonis 4572B} PMT is shown in Fig.~\ref{fig:Photonis_b&a}. The WLS layer can be removed from the PMT using a razor and acetone.

\section{`Table-Top' Setup to Test Different WLS Paints}
\label{sec:TTS}
\begin{figure}[htb]
\centering
\includegraphics[width=0.3\textwidth]{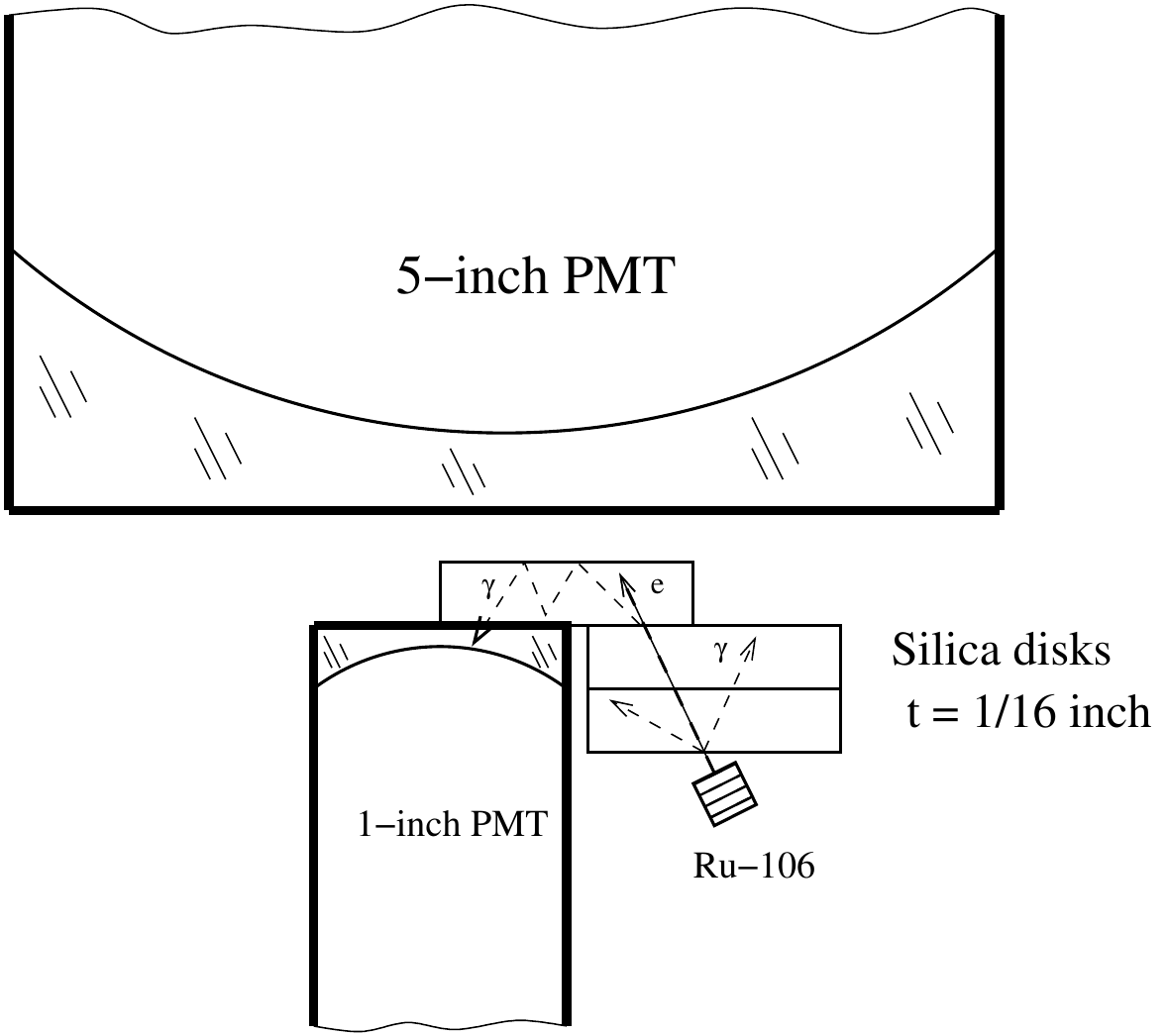}
\caption{The diagram of the `table-top' experimental setup. Dashed lines indicate the paths of Cherenkov photons. The 5-inch PMT was placed as close as possible to the silica disks. So only a small portion of the PMT surface was exposed to Cherenkov photons at one time, and the PMT had to be rotated about its axis to test other portions of the window.}
\label{fig:tt_diagram}
\end{figure}

The effects of the four different WLS paints were tested using {ET 9390KB} PMTs on a `table-top` setup. A diagram of the detector setup is shown in Fig.~\ref{fig:tt_diagram}. Collimated electrons from the $^{106}$Ru source produce Cherenkov radiation in the three fused silica disks. Cherenkov light passes through the disks and is detected by the {ET 9390KB} 5-inch PMT. Some of the light, though, is internally reflected inside the disks and detected by the 1-inch PMT, which is optically coupled to one of the disks. While allowing the signal from the 1-inch PMT to serve as the trigger, a full charge spectrum of the {ET 9390KB} PMT was observed. The trigger rate used was approximately 5-7 Hz, while the rate of coincidence between the two PMTs was 3-4 Hz. The accidental coincidence rate was found to be 0.01-0.02 Hz.

The equation used to calculate the average number of photoelectrons produced is $N_{PE} = \frac{A_{spec}}{mS-PE} + 0.5$, where $A_{spec}$ is the average of the charge spectrum (excluding the pedestal region) and mS-PE is the average of the S-PE spectrum. The addition of 0.5 photoelectrons is included because the Poisson distribution is treated as continuous in this calculation~\cite{TN}.

The maximum gain for the four different paints on a specific PMT is shown in Table \ref{table:wls}. As discussed in the previous section, the paints {EJ-298\#2} and {EJ-299-31E} could be applied to the PMT surface uniformly; they showed consistent gains across the whole surface of the PMT and when tested on other PMTs of the same type. The paints EJ-299-31A and EJ-299-31C gave a large signal increase at times (as shown in Table \ref{table:wls}), but the results varied significantly over the full surface of the PMT. From these considerations, we selected the paint {EJ-299-31E} for further tests using the electron beam in Hall A.

\begin{table}[htb]
\caption{Maximum signal gain on {ET 9390KB} PMT (SN:17707). The paints EJ-298\#2 and EJ-299-31E gave consistent results over the full surface of the PMT and when tested on other PMTs of the same type.}
\centering
\begin{tabular}{| c | c | c|}
\hline
WLS applied & $N_{PE}$ & Signal increase \\ \hline 
\hline
None         &  6.67    &  -    \\ \hline
{EJ-298\#2}  &  8.00    &  20\% \\ \hline 
{EJ-299-31A} &  9.43    &  41\% \\ \hline
{EJ-299-31C} &  9.18    &  37\% \\ \hline
{EJ-299-31E} &  8.94    &  34\% \\ \hline
\end{tabular}
\label{table:wls}
\end{table}

\section{Beam Tests of WLS Paint in Hall A at Jefferson Lab}

Hall A at Jefferson Lab is equipped with two high-resolution spectrometers (HRS), which detect charged particles at selected momenta~\cite{halla}. Each HRS has a Gas Cherenkov detector, which is used for electron identification. The Gas Cherenkov detectors are about 1.2~m in length and filled with CO$_2$ at atmospheric pressure. At the back of each Gas Cherenkov detector are 10 concave mirrors, which focus light onto a specific PMT~\cite{halla_gc}. 
The PMTs used in the Left-HRS Gas Cherenkov are {ET 9390KB} PMTs,
while {Photonis 4572B} PMTs are used in the Right-HRS spectrometer. 
The effect of the WLS paint {EJ-299-31E} was analyzed using GeV electrons detected in each HRS. The paint was tested on two PMTs on the Left-HRS and three PMTs on the Right-HRS. The other PMTs were used to check the stability of the setup.

\begin{figure}[htb]
\centering
\includegraphics[width=0.45\textwidth]{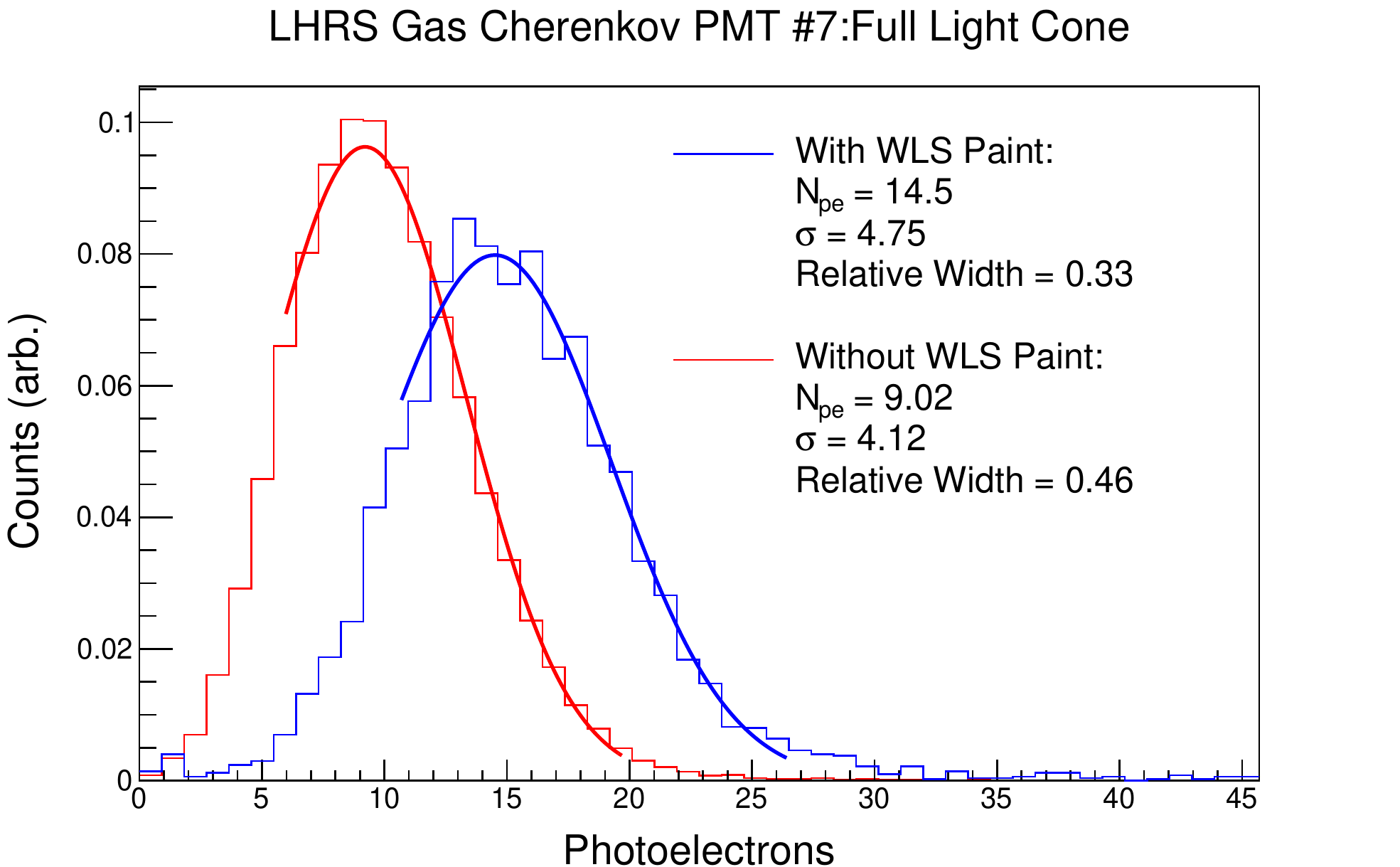}
\caption{Charge spectra for an {ET 9390KB} PMT on the Left-HRS.
The number of photoelectrons is calculated by the method used in the previous section. The improvement in the photon detection efficiency with the WLS paint is 61\%.
The relative width is defined as the rms ($\sigma$) divided by the mean of the Gaussian fit.} 
\label{fig:L_7}
\end{figure}

\begin{figure}[htb]
\centering
\includegraphics[width=0.45\textwidth]{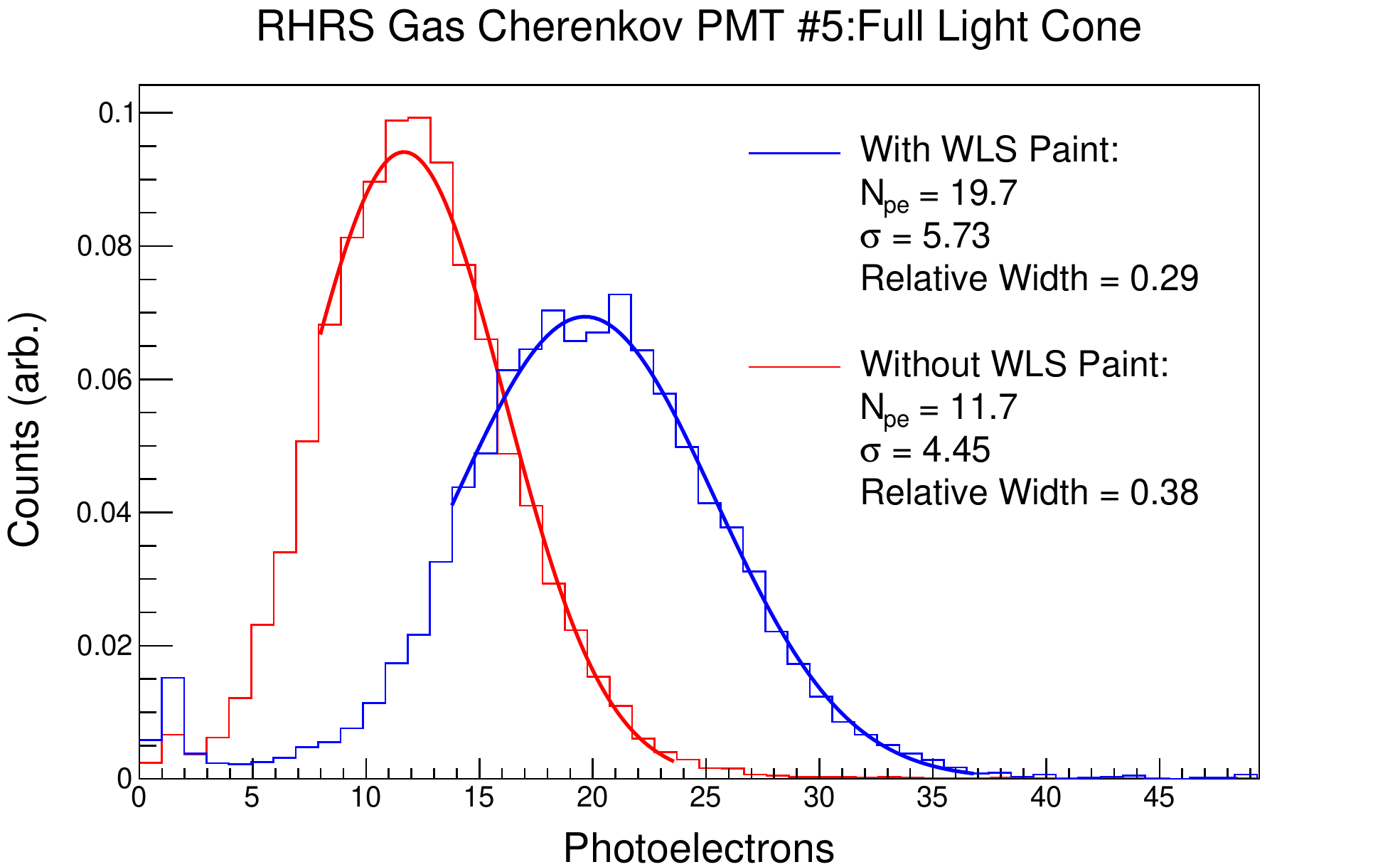}
\caption{Charge Spectra for a {Photonis 4572B} PMT on the Right-HRS. The improvement in the photon detection efficiency with the WLS paint is 68\%. } 
\label{fig:R_5}
\end{figure}

We selected events for which the Cherenkov radiation cone is fully contained within the corresponding mirror. The spectrum after applying this cut is shown in Fig.~\ref{fig:L_7}. It should be noted that the spectra is wider when the WLS paint is applied. This is to be expected because, as the number of photoelectrons produced increases, the number of S-PE spectra that need to be convoluted increases as well, so the spectrum will grow wider. The relative width of the distribution decreases, however, in accordance with Poisson statistics. The result for a {Photonis 4572B} PMT on the Right-HRS is shown in Fig.~\ref{fig:R_5}. 

A 65\% improvement in the number of photoelectrons produced was seen in both types of PMTs. The WLS paint's effect on the time resolution of the detector was found to be less than 2~ns. No degradation in the paint's performance has been found over the course of 6 months.

\section{Conclusion}
We presented a successful implementation of an inexpensive way to enhance the sensitivity of non-UV-transparent window based PMTs using WLS paint. The results from the electron beam tests at Jefferson Lab demonstrated a 65\% increase in the number of photoelectrons produced by {ET 9390KB} and {Photonis 4572B} PMTs for the WLS paint {EJ-299-31E}. 

Using the same WLS paint with the discussed `table-top' setup, however, only a 30-40\% increase was observed. This is thought to indicate that the trigger was not as clean as in the case of the GeV electron test.

\section*{Acknowledgments}
The authors would like to thank V.~Sulkosky and Y.~Wang for their contributions to the PMT testing, and S.~Gilad and C.~Keppel for their support of the project. This work was supported by the Department of Energy (DOE) contract number DE-AC05-06OR23177, under which the Jefferson Science Associates operates the Thomas Jefferson National Accelerator Facility.


\section*{References}
\begin{thebibliography}{99}

\bibitem{garwin} E.L.~Garwin, Y.~Tomkiewicz, and D.~Trines. {Nuclear Instruments and Methods} {\bf 107} (1973) 365.

\bibitem{baillon} P.~Baillon {\it et al.} {Nuclear Instruments and Methods} {\bf 126} (1975) 13.  

\bibitem{paneque} D.~Paneque {\it et al.} {Nuclear Instruments and Methods A} {\bf 504} (2003) 109.

\bibitem{CBM} J.~Adamczewski-Musch {\it et al.} {Nuclear Instruments and Methods A} {\bf 766} (2014) 180.

\bibitem{boccone} V.~Boccone {\it et al.} (The ArDM Collaboration) 2009 JINST {\bf 4} P06001.

\bibitem{hohne} C.~Hohne {\it et al.} {Nuclear Instruments and Methods A} {\bf 639} (2011) 294.

\bibitem{TUNKA}  N.~Surin {\it et al.} {Nuclear Instruments and Methods A} {\bf 766} (2014) 160.

\bibitem{eljen} Eljen Technology, \url{http://www.eljentechnology.com/index.php/products/paints}

\bibitem{TN} M.~Shepherd and A.~Pope, Investigation of BURLE 8854 
Photomultiplier Tube, JLab Technical Note 97-028. \url{http://hallaweb.jlab.org/12GeV/experiment/E12-07-108/Publications/Technical/Burle8854_technote.pdf}

\bibitem{NIM-D} R.~Dossi {\it et al.} {Nuclear Instruments and Methods A} {\bf 451} (2000) 623.

\bibitem{halla} J.~Alcorn {\it et al.} {Nuclear Instruments and Methods A} {\bf 522} (2004) 294.

\bibitem{halla_gc} M.~Iodice {\it et al.} {Nuclear Instruments and Methods A} {\bf 411} (1998) 223.

\end{thebibliography}
\end{document}